\documentclass{article}
\pdfoutput=1

\usepackage{arxiv}

\usepackage[utf8]{inputenc} 
\usepackage[T1]{fontenc}    
\usepackage[hidelinks]{hyperref}       
\usepackage{url}            
\usepackage{booktabs}       
\usepackage{amsfonts}       
\usepackage{nicefrac}       
\usepackage{microtype}      
\usepackage{lipsum}		
\usepackage{graphicx}
\usepackage{doi}
\usepackage{listings} 
\usepackage{tikz}
\usetikzlibrary{arrows} 
\usetikzlibrary{positioning, calc, fit}
\usetikzlibrary{shapes}
\usetikzlibrary{decorations.markings}
\tikzstyle{vecArrow} = [thick, decoration={markings,mark=at position
   1 with {\arrow[semithick, fill=black]{triangle 60}}},
   double distance=1.4pt, shorten >= 5.5pt,
   preaction = {decorate},
   postaction = {draw,line width=1.4pt, shorten >= 4.5pt}]

\newcommand{\codefont}{\scriptsize\sffamily} 
\newcommand{\codefontinline}{\small\sffamily} 

\lstdefinestyle{prg} {basicstyle=\codefont, lineskip=-0.2ex, showspaces=false}
\lstdefinestyle{prginline} {basicstyle=\codefontinline, lineskip=-0.2ex, showspaces=false}

\newcommand{\progcpp}[4][htb]{ 
 \begin{figure}[#1]
   \lstinputlisting[language=C++,style=prg,numbers=left,xleftmargin=2em,frame=single,framexleftmargin=2em]{figs/#2.cc}
  \caption[#4]{#3\label{fig:#2}}
 \end{figure}
}

\title{MARS: Middleware for Adaptive Reflective Computer Systems}

\author{Tiago M{\"u}ck \qquad Bryan Donyanavard \qquad \href{https://orcid.org/0000-0002-5830-861X}{\includegraphics[scale=0.06]{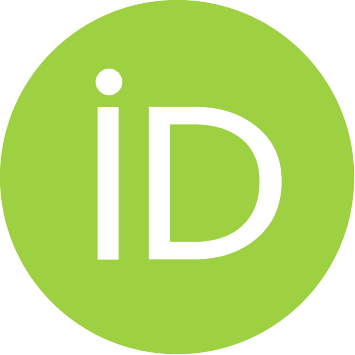}\hspace{1mm}Biswadip Maity} \qquad Kasra Moazzemi \qquad Nikil Dutt \\ 
Department of Computer Science \\
University of California, Irvine, USA \\
\texttt{\{tmuck,bdonyana,maityb,moazzemi,dutt\}@uci.edu} 
}

\renewcommand{\shorttitle}{MARS: Middleware for Adaptive Reflective Computer Systems}

\hypersetup{
pdftitle={MARS: Middleware for Adaptive Reflective Computer Systems},
pdfsubject={cs.AR},
pdfauthor={Biswadip Maity},
pdfkeywords={Artifact, Framework, Reflection, Adaptive, Middleware, Runtime resource management},
}

\begin{document}
\maketitle

\begin{abstract}

Self-adaptive approaches for runtime resource management of manycore computing platforms often require a runtime model of the system that represents the software organization or the architecture of the target platform.
The increasing heterogeneity in a platform's resource types and the interactions between resources pose challenges for coordinated model-based decision making in the face of dynamic workloads. 
Self-awareness properties address these challenges for emerging heterogeneous manycore processing (HMP) platforms through reflective resource managers. 
However, with HMP computing platform architectures evolving rapidly, porting the self-aware decision logic across different hardware platforms is challenging, requiring resource managers to update their models and platform-specific interfaces. 
We propose MARS (Middleware for Adaptive and Reflective Systems), a cross-layer and multi-platform framework that allows users to easily create resource managers by composing system models and resource management policies in a flexible and coordinated manner.
MARS consists of a generic user-level sensing/actuation interface that allows for portable policy design, and a reflective system model used to coordinate multiple policies.
We demonstrate MARS' interaction across multiple layers of the system stack through a dynamic voltage and frequency scaling (DVFS) policy example which can run on any Linux-based HMP computing platform. 

\end{abstract}


\section{Introduction}

At the processing platform level, heterogeneous processing elements have become increasingly popular as a way to exploit the power-performance tradeoff for dynamic workloads. 
This growth in popularity can be observed across multiple domains, ranging from portable battery-powered devices to supercomputers. 
Emerging mobile battery-powered Systems-on-Chip (SoCs) incorporate heterogeneity in order to provide energy-efficiency. 
For instance, ARM's \emph{big.LITTLE} architecture \cite{ARM2013c} is used on Samsung's Exynos \cite{Exynos2016} and Nvidia's Tegra SoCs \cite{NVidia2011b}. 
big.LITTLE employs \emph{single-ISA heterogeneity}, in which cores of the same instruction-set architecture (ISA) are deployed with different configurations and/or microarchitectures (for example, in-order cores operating at low frequency \textit{vs.} out-of-order cores operating at high frequency).

Effective exploitation of power-performance tradeoffs in heterogeneous manycore processors (HMPs) requires intelligent management both within and across different layers, in particular the operating system \cite{Rahmani2017}.
Operating systems must continuously analyze application behavior to resolve tradeoffs: e.g., \emph{What is the most power efficient core type to execute the application without violating its performance requirements?}
or \emph{Which option is more power-efficient for the current application: an out-of-order core at a lower frequency or an in-order core at a higher frequency?}

Being able to properly assess the performance/power impact of resource management decisions plays a major role in efficiently managing the system, especially on HMPs in which different resources (e.g., CPU cores, memory buses, GPU cores, etc.) have distinct performance/power tradeoffs.
Some existing Linux extensions address these issues to a limited extent.
For instance, Global Task Scheduling (GTS) \cite{ARM2013c} migrates tasks between high-performance and low-power core types when the task load reaches a certain threshold.
GTS-based policies have been implemented on Linux and deployed by multiple SoC vendors that support the big.LITTLE technology (e.g., Linaro's/ARM's big.LITTLE MP implementation used by Samsung Exynos \cite{Poirier2014}, MediaTek's CorePilot \cite{MediaTek}, Qualcomm's Energy Aware Scheduling \cite{Muckle2014}).
These policies, however, are customized for specific heterogeneous platforms without being adaptable to other platforms.
Furthermore, they exclusively focus on task mapping/scheduling decisions and still do not provide any coordination with the  underlying power management subsystems.

State-of-the art works address these issues by applying principles of self-awareness.
Self-aware computing systems are computing systems that capture knowledge about themselves and their environment on an ongoing basis to learn models, and reason about the models in order to make decisions in accordance with high-level goals, which are subject to change \cite{Kounev2017}.
The notion of self-awareness has been used in computing in a variety of different domains such as autonomic computing, organic computing, adaptive systems, and self-organizing systems \cite{jantsch2017self}.
Although adaptive systems with no self-awareness exist, it has been argued that a sophisticated self-model is a prerequisite for sensible adaptive behavior when the environment and the system itself are sufficiently complex \cite{Salehie:2005:ACE:1082983.1083082}.

Some apply self-reflection by using models that are able to estimate or predict workload behavior by extracting 
performance metrics at runtime~\cite{SomuMuthukaruppan2014,VanCraeynest2012,Sarma2015}.
While a significant amount of work has been done on the management of specific resources (e.g. DVFS\cite{Nejatollahi2015,Shafik2016,Donyanavard2018b},
task mapping\cite{Paterna2013a,Sahoo2016,Trainiti2016}, memory 
allocation\cite{Muck2011,Shoushtari2015,Shoushtari2018}), most do not provide means for coordinating the management of different resource types.

Self-adaptive software systems and applications have been successfully deployed \cite{HALLSTEINSEN2012, Geihs2009, gomaa2004, Morin:2009:TDA:1555001.1555028}, 
for example to extract information from networks of cameras \cite{gurgen2013}.
Hoffman et al. developed the SEEC framework \cite{hoffmann2010seec} for self-aware \emph{applications}. Leech et al. developed the PRiME Framework \cite{DBLP:conf/recosoc/LeechBBW18} to provide support for testing resource managers across platforms.
Our goal is to similarly provide a framework for \emph{policy} designers to design adaptive resource managers that support properties of self-awareness. 

In this artifact paper, we present MARS  (Middleware  for  Adaptive  and Reflective Systems) \cite{drg_mars},
a cross-layer and multi-platform framework supporting the creation of resource managers for emerging  heterogeneous manycore  processing  (HMP)  platforms by composing system  models  and  resource  management  policies  in  a  \textbf{flexible} and \textbf{coordinated manner}. 
MARS has been utilized in different configurations for research projects such as SPARTA \cite{Donyanavard2016}, SPECTR \cite{Rahmani2018}, and deployed on multiple hardware platforms.
Users define policies using generic interfaces so that they are \textbf{portable} across any supported hardware platform.
The modular composition makes the framework easily \textbf{expandable} to support new platforms, sensors, or actuators.
MARS interacts with all layers of the system stack to orchestrate the management of resources.
MARS is mainly composed of four parts:
\begin{enumerate}
 \item \textbf{Sensors and actuators.} The sensed data that consists of performance counters (e.g. instructions executed, cache misses, etc.) and other sensory information (e.g. power, temperature, etc.) collected to assess the current system state and to characterize workloads. The configurable knobs that allow the platform to adjust its configuration to optimize operating point or control tradeoffs.
 \item \textbf{Resource management policies} implemented by MARS's users.
 \item \textbf{Reflective system model} used by the policies to make informed decisions. The reflective model has the following subcomponents:
 \begin{enumerate}
  \item Models of \emph{policies implemented by the underlying OS kernel} used for coordinating decisions made within MARS with decisions made by the OS.
  \item Models of \emph{user policies}, that are automatically instantiated from any policy defined within MARS.
  \item The baseline \emph{performance/power model}. This model takes as input the predicted actuations generated from the policy models and produces predicted sensed data.
 \end{enumerate}
 \item \textbf{The policy manager} is responsible for reconfiguring the system by adding, removing, or swapping policies to better achieve the current system goal.
\end{enumerate}

\section{Background: Runtime Models, Reflection and Prediction}

Self-adaptive software can be defined as ``software that evaluates its own behavior and changes behavior
when the evaluation indicates that it is not accomplishing what the software is intended
to do, or when better functionality or performance is possible''\cite{Laddaga2000}.
In this work, we do not address self-adaptive systems directly, as these systems encompass many high-level concepts such as self-configuration, self-healing, self-awareness, self-optimization, and others, also referred to as \emph{self-$*$}\cite{Salehie2009} (see \cite{Dutt2016} for a comprehensive review self-adaptivity for SoCs). 
However, an infrastructure for system introspection and reflective behavior is an important building block for such systems. 

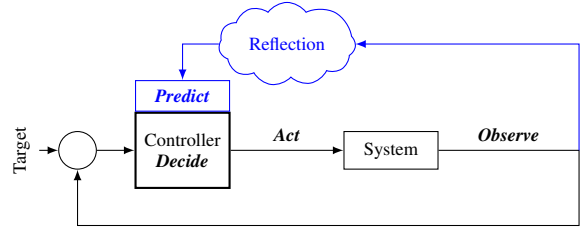
\begin{figure}
    \centering
        \begin{tikzpicture}[font={\scriptsize}, >=latex, every text node part/.style={align=center}]
	\def\mwdth{1.25cm}
	\def\mhgth{0.5cm}
	\def\mdis{0.5cm}
	\node [draw, minimum width=\mwdth, minimum height=\mhgth*2, thick] (Controller) at (0,0) {Controller\\\textit{\textbf{Decide}}};
	\node [draw, color = blue, minimum width=\mwdth, minimum height=\mhgth*0.5, above = 0 of Controller] (Prediction) {\textit{\textbf{Predict}}};
	\node [draw, minimum width=\mwdth, minimum height=\mhgth, right = 3*\mdis of Controller] (System) {System};
	\node [circle, draw, minimum width=1*\mhgth, minimum height=1*\mhgth, left = \mdis of Controller] (Diff) {};
	\draw[->] (Diff) -- (Controller);
	\draw[->] (Controller) -- (System) node [midway, above] (Act_Freq) {\textit{\textbf{Act}}} ;
	\node [left= 0.1*\mdis of Diff, opacity=0.0] (tmp) {Target};
	\node [rotate around={90:(tmp.center)}] (Ref) at (tmp) {Target};
	\draw[->] (Ref) -- (Diff);
	\coordinate[] (FBhelp) at ($(System) + (5*\mdis, -2*\mdis)$);
	\draw[->] (System) -| node [near start, above] (Sen_Heart) {\textit{\textbf{Observe}}} node [near start, below] (Sen_IPS) {} (FBhelp) -| (Diff);
	\node[cloud, cloud puffs=10.8,cloud puff arc=110, aspect=2, draw, color = blue, above = \mdis of Act_Freq 
    ] (Reflection)  {Reflection};
	\coordinate[] (Refhelp) at ($(System) + (5*\mdis, 0)$);
    \draw[->, color=blue] (Refhelp) |- (Reflection);
    \draw[->, color=blue] (Reflection) -| (Prediction);
\end{tikzpicture}
    \caption{Feedback loop overview. The bottom part of the figure represents a simple observe-decide-act loop. The top part (in \textcolor{blue}{blue}) adds the reflection mechanism to this loop, enabling predictions for smart decision making.}
    \vspace{-0.2in}
    \label{fig:FB-Controller}
\end{figure}


\begin{figure*}[ht]
\centering
        \includegraphics[width=0.9\textwidth]{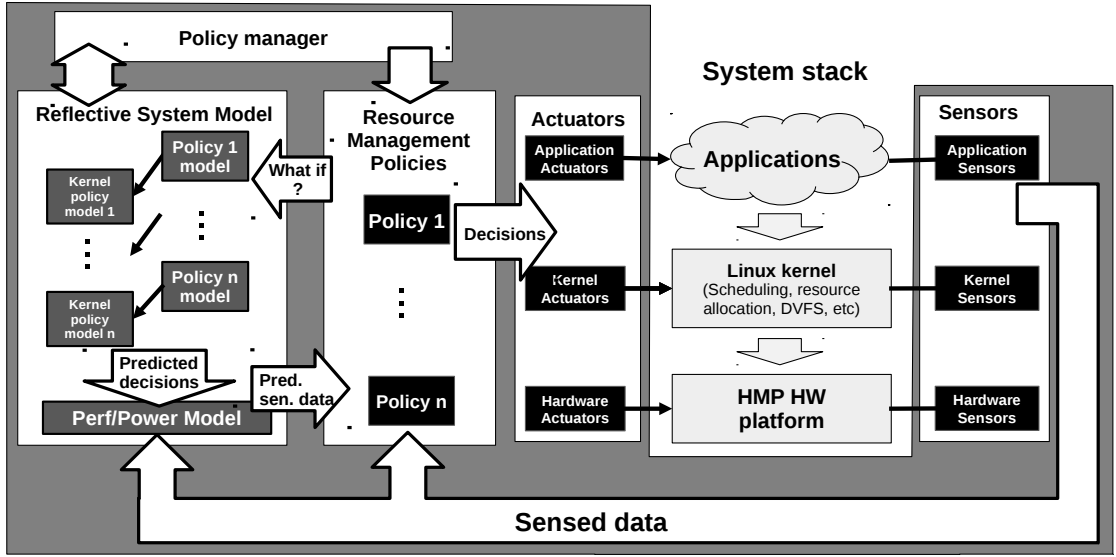}
        \caption{MARS framework overview. Different layers of the system stack coordinate through policies to orchestrate the management of resources: sensors inform policies of the system state; policies coordinate with models to perform reflective queries, and make resource management decisions; policies set actuators to enact changes on the system. Details about sensors and actuators in Table \ref{tab:sensors_actuators}.}
        \label{fig:framework}
\end{figure*}

\subsection{Observe, Decide, and Act, plus Reflection}

Closed-loop systems have been used extensively to improve the state of a system by tracking target references. 
Closed-loop systems traditionally deploy an \emph{Observe}, \emph{Decide} and \emph{Act} (ODA) feedback loop (lower half (in black) of Figure~\ref{fig:FB-Controller}) to determine the system configuration.
In an ODA loop, the observed behavior of the system is compared to the target behavior, and the discrepancy is fed to the controller for decision making. 
The controller invokes actions based on the result of the \emph{Decide} stage. 
MARS enables an additional \emph{Reflection} loop (upper half (in \textcolor{blue}{blue}) of Figure~\ref{fig:FB-Controller}) that incorporates the ability to include past history, as well as predictions, to make smart decisions and enhance adaptivity.





Reflection is a key property of self-awareness.
Reflection enables ODA decisions to be made based on both \emph{past} observations, as well as predictions made from past observations.
MARS ODA with reflection and prediction requires a self-model of the subsystem(s) under control, as well as models of other policies that may impact the decision-making process.
MARS predictions consider \emph{future} actions, or events that may occur before the next iteration of the ODA loop, enabling "what-if" exploration of alternatives. 
Such actions may be triggered by other resource managers running with a shorter period than the ODA loop. 
The top half of Figure~\ref{fig:FB-Controller} (in \textcolor{blue}{blue}) shows prediction enabled through reflection that can be utilized in the decision making process, as supported by MARS.
The main goal of the prediction model is to estimate system behavior based on potential actuation decisions. 
This type of prediction is most often performed using linear regression-based models \cite{Muck2015,Pricopi2013} due to their 
simplicity, while others employ a binning-based approach in which metrics sensed at runtime are used to classify workloads into categories whose performance/power are known for all core types\cite{Liu2013,Donyanavard2016}.

\section{MARS Framework}

Figure \ref{fig:framework} shows an  overview of the MARS framework (shaded), with 
\emph{Sensors}
and 
\emph{Actuators}
interfacing across multiple layers of the system stack: 
\emph{Applications},
\emph{Linux kernel}, and
\emph{HMP HW Platform}. 
On-chip resource management requires modeling and runtime policies for these different layers of the system stack.
Resource management decisions must be made for each layer, and the frequency at which decisions are made may vary within or between layers. 
Coordinating multiple resource management policies is challenging, and that challenge compounds when developers are required to design portable policies for multiple platforms.
We have developed MARS -- 
\emph{Middleware for Adaptive and Reflective Systems} -- to address these challenges. 

\subsection{Interfaces and Policy Design}
We first look into how MARS provides a generic user-level sensing/actuation interface that allows for \textbf{portable} policy design. Table \ref{tab:sensors_actuators} provides a list of common sensors and actuators available at each level of the system stack. 

\subsubsection{Sensors}
Sensor data may consist of physical or virtual performance counters (e.g., instructions executed, cache hits), or other sensory information (e.g., power, temperature). 
Sensor data assesses the current state of the system and also characterizes the workload. 
All sensors expose a \emph{virtual interface} for communication with the resource manager. 
To implement the interface, MARS utilizes function templates to provide a generic \texttt{sense} function that returns the specified sensed metric.
Different versions of the function can be implemented for each platform, resource, and sensor-type combination.
The function's template parameter specifies the proper function implementation to obtain the requested sensed data.
This separates the platform-specific implementation from the user-level policies.


\begin{table}[tbh]
\centering
\begin{tabular}{p{\dimexpr 0.18\linewidth-2\tabcolsep} 
            p{\dimexpr 0.35\linewidth-2\tabcolsep} 
            p{\dimexpr 0.35\linewidth-2\tabcolsep}}
    \toprule
    Level & Actuators (\textbf{A}) & Sensors (\textbf{S})\\
    \midrule
    Application &
    Degree of parallelism, algorithmic choice &
    Performance, Quality-of-Service\\ 
    \midrule
    Kernel &
    Task-to-core mapping, task priority, memory allocation &
    CPU time, utilization, context switch counters, open file, number of tasks \\
    \midrule
    Hardware &
    Voltage/frequency scaling, clock gating, power gating, cache sizing  &
    Performance counters (e.g., cache misses, instructions executed, branch miss-predictions, etc), power, reliability, critical path delay degradation \\
    \bottomrule
\end{tabular}
\vspace{10pt}
\caption{Examples of sensors and actuators available across the system stack}
\vspace{-0.2in}
\label{tab:sensors_actuators}
\end{table}

\subsubsection{Actuators}
Adjustments to system configuration at runtime happens through actuators. 
Actuators range from application-level choices (e.g., degree of parallelism), to kernel-level choices (e.g., task-to-core mapping), to hardware-level choices (e.g., core operating frequency).
The \texttt{actuate} function sets a new value for a system actuator using a \emph{virtual interface}. 
Similarly to \texttt{sense}, \texttt{actuate} is implemented with function templates so that unique versions may be provided for each platform, resource, and actuator combination.
Given the new actuation value, the function's template parameter selects the proper function implementation to perform the requested actuation action. 
Two additional related functions are provided: \texttt{actuationVal} returns the current actuation value set and \texttt{actuationRanges} returns the valid range of actuation values for the specified resource.
\texttt{actuate} functions are implemented on top of standard Linux system calls and modules.
For instance, we use CPU affinity system calls to control task-to-core mappings, and \emph{cpufreq} to control core frequencies.

\subsubsection{Policies}
\label{sec:policy}
Policies contain the decision-making logic which controls the system resources at runtime.
Management policies implement the ODA and refelction loops for the resource(s) under control.
In MARS, users can define resource management policies by creating a subclass of \texttt{Policy}. 

All policies make use of generic sensing and actuation interfaces \texttt{sense} and \texttt{actuate}. 
At the user-level, all policies share access to the common set of sensors. 
For each policy, sensor data is aggregated during the \textit{sensing window} (i.e., the time period between policy invocations) defined for the policy.
MARS uses \texttt{ioctl} system calls to setup the sensing windows for each registered policy function.
The policy functions are executed at the end of each sensing window in separate threads that uses blocking \texttt{ioctl} calls to synchronize with the kernel module.
The kernel module automatically aggregates sensed data on a per-window basis and stores the information in shared memory, so that it can be read directly by \texttt{sense} calls with low overhead.
This means any number of sensing windows (i.e., policies) executing at different window lengths can access the same sensors concurrently, and still measure the appropriate value for their window length.

\subsection{Reflective System Model}
Previously, we described a scenario where multiple policies operate at different sensing window lengths, or epochs. 
We  now illustrate how MARS uses a reflective system model to make self-aware decisions. 





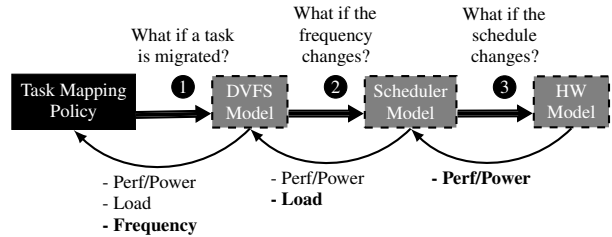
\begin{figure}[tbh]
\centering
        \begin{tikzpicture}[scale=0.6, node distance = 1.5cm, font={\scriptsize}, >=latex, every text node part/.style={align=center}]
	\def\mwdth{1cm}
	\def\mdis{0.5cm}
	
	\node [draw, fill=black, text=white, minimum width=\mwdth, minimum height=0.5cm, ] (TM) {Task Mapping\\Policy};
	\node [draw, fill=gray, text=white, thick, dashed, minimum width=\mwdth, minimum height=0.5cm, right = 2*\mdis of TM] (DVFS) {DVFS\\Model};
	\node [draw, fill=gray, text=white, thick, dashed, minimum width=\mwdth, minimum height=0.5cm, right = 2*\mdis of DVFS] (SCHED) {Scheduler\\Model};
	\node [draw, fill=gray, text=white, thick, dashed, minimum width=\mwdth, minimum height=0.5cm, right = 2*\mdis of SCHED] (HW) {HW\\Model};
	
	
	\coordinate[below left = -2mm and 0mm of DVFS] (help0);
	\coordinate[below right = -2mm and 0mm of TM] (help1);
    \draw [vecArrow] (help1) -- (help0);
	\node [above left = 0cm and -0.5cm of DVFS] (ctrl_params) {\scriptsize \begin{parbox}{1.5cm}{\centering What if a task is migrated?}\end{parbox}};
	\node [draw, circle, fill=black, text=white, minimum size = 0.3cm, inner sep=0pt, below = 0 of ctrl_params] (label) {1};
	\coordinate[below left = -2mm and 0mm of SCHED] (help0);
	\coordinate[below right = -2mm and 0mm of DVFS] (help1);
    \draw [vecArrow] (help1) -- (help0);
	\node [above left = 0cm and -0.5cm of SCHED] (ctrl_params) {\scriptsize \begin{parbox}{1.5cm}{\centering What if the frequency changes?}\end{parbox}};
	\node [draw, circle, fill=black , text=white, minimum size = 0.3cm, inner sep=0pt, below = 0 of ctrl_params] (label) {2};
	\coordinate[below left = -2mm and 0mm of HW] (help0);
	\coordinate[below right = -2mm and 0mm of SCHED ] (help1);
    \draw [vecArrow] (help1) -- (help0);
	\node [above left = 0cm and -0.5cm of HW] (ctrl_params) {\scriptsize \begin{parbox}{1.5cm}{\centering What if the schedule changes?}\end{parbox}};
	\node [draw, circle, fill=black, text=white, minimum size = 0.3cm, inner sep=0pt, below = 0 of ctrl_params] (label) {3};

    \draw[->, bend left=45, thick] (DVFS.south) to node[auto] {\begin{parbox}{1.5cm}{- Perf/Power\\- Load\\\textbf{- Frequency}}\end{parbox}}(TM.south);	
    \draw[->, bend left=45, thick] (SCHED.south) to node[auto] {\begin{parbox}{1.5cm}{- Perf/Power\\\textbf{- Load}}\end{parbox}}(DVFS.south);	
    \draw[->, bend left=45, thick] (HW.south) to node[auto] {\begin{parbox}{1.5cm}{\textbf{- Perf/Power}}\end{parbox}}(SCHED.south);	
    
	
\end{tikzpicture}
        \caption{Example of a task mapping policy that queries models of OS policies for DVFS and scheduling.}
        \vspace{-0.2in}
        \label{fig:framework_example_on_framework}
\end{figure}

To understand the reflection mechanism in MARS, consider the resource manager shown in Figure \ref{fig:framework_example_on_framework} 
that shows a sample task mapping policy interacting with DVFS (Dynamic Voltage and Frequency Scaling) and scheduler policy models. 
At the finest time granularity, we have the operating system scheduler, whose goal is to select a task to execute on a given core. 
A new decision must be made whenever a new task is created, a tasks finishes, a task's quantum expires, an interrupt is raised, etc., leading to a time-frame between decisions on the order of microseconds. 
At a coarser time granularity we have the DVFS policy, which typically executes periodically (10-100 milliseconds) to analyze the system load and select an appropriate operating frequency. 
At the coarsest time granularity (100-1000 of milliseconds), the task mapping policy runs periodically to define a new task-to-core assignment. 
Migrating a task from one core to another has significantly more overhead than changing the CPU frequency in a typical HMP \cite{Kisoo2012}. 
MARS allows users to \textbf{coordinate} among different policies through policy models, regardless of varying time granularity.

In order to make an informed task mapping decision, for instance, the policy must consider the effects of its decision on the behavior of the underlying DVFS and scheduling policies. 
Furthermore, the invocation period of actuations dictates how complex the decision making logic can be. 
For instance, a scheduling decision must be made in the sub-microsecond range in order not to disrupt the system. 
Task-to-core mapping, on the other hand, is done comparatively infrequently, and affects the system performance over a long timespan.
Therefore, the overhead of using complex models to make such decisions can be mitigated by the potential benefits of an informed decision.

The components within the reflective system model interact in a hierarchy defined by the dependencies of the actuations performed in the system.
For instance, Figure \ref{fig:framework_example_on_framework} illustrates the scenario within MARS for our example. 
\textbf{Workload models} assume each core can run multiple tasks and there is no formal or explicit dependency between threads. 
Before the task mapping policy decides to migrate a task, it (1) queries the reflective model asking: \emph{what will be the performance of task A if it is migrated?} 
(2) The DVFS governor policy model executes (\emph{without} actuating) in order to predict the resulting core frequency provided the hypothetical task mapping. 
(3) This information is passed on to the performance/power model which predicts the task performance. 
\textbf{Architecture models} define the architectural characteristics of the target platform including \emph{instruction-set architecture} (ISA), number of cores, core types, etc. 
Finally, the predicted metrics are used by the policy to make the decision, which is passed to the actuator through the actuation interface.

Analogous to the \texttt{actuate}/\texttt{sense} functions described previously, MARS provides the \texttt{tryActuate} and \texttt{senseIf} set of functions necessary issuing queries to the reflective models:
\begin{itemize}
    \item \texttt{tryActuate}: Using the same syntax as \texttt{actuate}, this function updates the underlying models used to predict the next system state given the new actuation value. It does not set the actual actuation value. A \texttt{tryActuationVal} analogous to \texttt{actuationVal} is also provided.
    \item \texttt{senseIf}: this function has the same semantics as \texttt{sense}, but returns predicted sensed information for the next sensing window, given a new actuation set by \texttt{tryActuate}.
\end{itemize}


\subsection{Policy Manager}
A user-defined subclass of \texttt{PolicyManager} is responsible for composing policies and models together. 
The purpose of the policy manager is to provide resource management autonomy in response to changing system goals or execution scenarios, as well as coordinate multiple dependent policies and their objectives. 
For instance, consider modern smartphones. 
These devices typically operate in two scenarios: 1) the device is plugged to an external power source; 2) the device is powered by battery. 
In the case of (1), policies can simply focus on maximizing applications' quality of service (QoS), while for (2), QoS should be balanced with energy efficiency. 
A third scenario in which the battery charge is critically low is also possible. 
In this case, policies should focus on minimizing power consumption so the device can continue to operate. 
Furthermore, direct intervention from the user may cause additional unpredictable scenario changes.
Creating a single policy that is able to manage all these scenarios and goals can lead to a overly complex and possibly inefficient implementation. 
Instead, one may develop multiple policies optimized for specific cases, and have a high-level manager dynamically  select the most appropriate policy to apply throughout runtime.


\begin{table*}[tb]
\caption{
Currently supported platforms in MARS and their sensors/actuators
}\label{tab:comparison}
\begin{center}
    {
    \begin{tabular}{p{\dimexpr 0.08\linewidth-2\tabcolsep} 
            p{\dimexpr 0.10\linewidth-2\tabcolsep} 
            p{\dimexpr 0.10\linewidth-2\tabcolsep} 
            p{\dimexpr 0.30\linewidth-2\tabcolsep} 
            p{\dimexpr 0.30\linewidth-2\tabcolsep}}
    \multicolumn{1}{c}{Implementation} &
    \multicolumn{1}{c}{ Platform } & 
    \multicolumn{1}{c}{ Architecture } &
    \multicolumn{1}{c}{ Sensors } &
    \multicolumn{1}{c}{ Actuators } 
    \\
    \toprule
    ODROID & Hardware & ARMv7-A & All ARM Performance Monitoring Unit (PMU) counters (e.g., cache misses, instructions executed, branch misspredictions, etc.,), Power sensors for CPU clusters & Voltage/frequency scaling for CPU, Task-to-core mapping, Temperature sensor
    \\ \midrule
    NVIDIA Jetson-TX2  & Hardware & ARMv8-A  & ARM PMU counters, GPU performance counters, CPU Power, Memory Power, GPU Power, WiFi Power &  Voltage/frequency scaling for CPU clusters, GPU frequency scaling, Task-to-core mapping mapping
    \\ \midrule
    Gem5 & Simulation & $*$  & ARM PMU counters & Frequency scaling
    \\ \midrule
    Offline \\simulator  & Simulation & $*$  & Sensors which are present in the trace & Switch across different traces of execution
    \\ \bottomrule
    \end{tabular}
    }
    \vspace{-0.2in}
\end{center}
\end{table*}

\subsection{DVFS Policy Example}
We first explain dynamic voltage and frequency scaling (DVFS) through a smartphone use case. 
When running multimedia/gaming applications, computational resource requirements are very high. 
To improve performance, CPU core frequency is increased dynamically, also resulting in higher power consumption. 
When the demand for computational resources is lowered, the CPU core frequency is reduced to save power. 
Using the MARS framework, we demonstrate a simple use case of a core DVFS policy. 

\progcpp{dvfs_example}
{Reflective DVFS management policy.}
{Reflective DVFS management policy.}
Our \texttt{Simple\_DVFS\_Policy} implemented in MARS is shown in Figure \ref{fig:dvfs_example}.
First, the \texttt{Simple\_DVFS\_Policy} is registered by the \texttt{Simple\_DVFS\_Manager} (\texttt{PolicyManager}) during \texttt{setup} using the \texttt{registerPolicy} function (lines 37-40). 
At this point user-defined models (if any) are also registered (using the \texttt{registerModel} function).
The DVFS policy is defined by the \texttt{Simple\_DVFS\_Policy}. 
\texttt{Simple\_DVFS\_Policy} inherits from \texttt{Policy} and implements its decision-making logic in the \texttt{execute} function (lines 16-35). 
When inheriting from \texttt{Policy} one must also define the policy invocation period (line 4). 
This period defines how often the policy is invoked as well as the hierarchy of the policies within the reflective model.
The period establishes a \emph{sensing window} for the policy. 
In this example, when the \texttt{execute} function is invoked, the sensed information (e.g., performance counters, power sensors, etc.) aggregated over the latest $50ms$ sensing window can be obtained.

Within the \texttt{execute} function, a special \texttt{sys\_info} object provided by the framework is used to access available resources in the system. 
The \texttt{sys\_info} also provides information about the topology and relationship between resources. 
Then, for all supported frequencies, the \texttt{try\_frequency} (Lines 6-14) function is called to assess the energy efficiency of the given frequency and select the best setting. 

Finally, the developer may use the same infrastructure to model the aspects implemented within the underlying system.
The \texttt{registerModel} function is provided to facilitate this process. 
When registered using \texttt{registerModel}, the policy code will only be executed as part of the model to predict performance and power, but it will not directly actuate on the system.


The above exemplar represents a naive DVFS governor for an HMP.
More sophisticated resource managers have been successfully deployed using MARS: \cite{Donyanavard2018b} deploys a gain-scheduled DVFS governor for dynamic power management; SPARTA \cite{Donyanavard2016} expands predictive power/performance model to perform runtime task allocation in HMPs; SPECTR \cite{Rahmani2018} is a hierarchical manager that coordinates an adaptive per-core clock gating coordinated with per-cluster DVFS for HMPs; \cite{Donyanavard2017} uses the offline-simulator to perform a design space exploration of hybrid on-chip memory configuration for HMPs; HESSLE-FREE \cite{10.1145/3358203} uses MARS for resource management by leveraging fuzzy control techniques. MARS has also successfully been deployed for runtime memory management techniques \cite{9291086,iess2019}. 

\section{State of Implementation}

MARS is implemented in the C++ language following an object-oriented paradigm. The framework is open source and available online.\footnote{Code repository at https://github.com/duttresearchgroup/MARS.}
In order to easily compile runtime resource management policies using MARS framework, a docker-based container image\footnote{Docker image at https://hub.docker.com/r/duttresearchgroup/mars.} is also available for developers.

\subsection{Platforms}
As highlighted in Table~\ref{tab:comparison},
MARS has been designed to work in conjunction with the Linux operating system and has been implemented on both hardware and simulation platforms. 
This enables policy evaluation on hypothetical system configurations \cite{Muck2017} in advance of policy deployment on a common hardware platform. 
We provide brief descriptions of these platforms below.

\subsubsection{ODROID}
Odroid-XU3\cite{Hardkernel2016} features a Samsung Exynos5422 HMP, which implements a 8-core big.LITTLE-based HMP. 
Cores are organized into clusters of the same type (i.e. a 4-core ``big'' cluster and a 4-core ``little'' cluster). 
Cores in the same cluster share the the same L2 cache and the same frequency domain. 
Odroid-XU3 also features per-cluster power sensors and temperature sensors for the GPU and for each core in the big cluster. 

\subsubsection{NVIDIA Jetson-TX2}
This NVIDIA Jetson-TX2 has a hex-core ARMv8 CPU (quad-core ARM Cortex-A57 alongside Denver cores) and a 256-core NVIDIA Pascal GPU. The high performance design and extensive support for AI and neural network application makes this board a great candidate to expand MARS to support new software domains. 

\subsubsection{gem5}

The gem5 simulator\cite{Binkert2011a} is a modular platform for computer architecture research, encompassing system-level architecture as well as processor microarchitecture.
We use gem5 to create heterogeneous cores by changing the architectural simulation parameters.
This allows us to evaluate MARS on HMP platforms with more than two core types.
For obtaining power data, we integrate the McPAT power model\cite{Li2013a} directly with the gem5 simulation framework, which allows us to obtain power data online and reduces simulation overhead. 
%
%
We use gem5 in full system mode which a complete operating system boots on top of gem5 and system calls are executed natively while supporting multithreaded workloads.



\subsubsection{Offline simulator}

Debugging and evaluating intelligent management policies on cutting-edge platforms is not straightforward.
Issues that are typically straightforward to fix in a prototype running as a user-space process can be highly problematic when they occur, for instance, within a module running in kernel-space.
Evaluating the new policy in terms of scalability and generality becomes much harder, since real platforms have fixed configurations
and full system simulations of large-scale systems are often unfeasible due to extremely long simulation times.
For such a scenario, we include a \emph{trace-based offline simulator} platform that supports MARS abstractions.
With the offline simulator and the Linux-based implementation, we can close the gap between proposing a new policy and implementing it on a real system. \cite{Muck2017} describes the offline simulator in more details.

\section{Conclusion and Future Work}
In this artifact paper, we presented  MARS:
a cross-layer, and multi-platform framework that supports coordinated self-adaptive resource management policies. 
Using MARS, policies are able to interact with one-another and make model-based decisions to adapt to dynamic workloads.
The MARS framework is open source and downloadable  as a  container-based docker image~\cite{drg_mars}.
While the current version of MARS targets energy-efficient heterogeneous multiprocessor (HMP) systems, we believe the MARS framework can be ported to a wider range of systems (e.g., webservers, high performing clusters), which opens up several directions for future work. 
As we continue to explore interaction among multiple policies, it has not escaped our attention that MARS currently does not have any validation mechanism for the emergent behavior of these policies acting together. 
In our future work, we plan to expand MARS to run on more complicated systems and develop a runtime validation mechanism for coordinated policies. 

\section*{Acknowledgment}
We acknowledge financial support from the following: National Science Foundation Grant CCF-1704859. We would also like to acknowledge the critical review by Saehanseul Yi on the manuscript.

\bibliographystyle{unsrt}
\bibliography{refs}  

\end{document}